\documentclass[twocolumn,showpacs,preprintnumbers]{revtex4}
\usepackage{psfig}
\begin{document}

\preprint{FZJ-IKP-TH-2006-13, HISKP-TH-06/13}

\title{
Near threshold  {\boldmath $p{\bar p}$} enhancement  in 
{\boldmath $B$} and {\boldmath$J/\Psi$} decay}

\author{J. Haidenbauer$^1$, Ulf-G. Mei{\ss}ner$^{1,2}$, A. Sibirtsev$^{2}$}

\affiliation{
$^1$Institut f\"ur Kernphysik (Theorie), Forschungszentrum J\"ulich,
D-52425 J\"ulich, Germany \\
$^2$Helmholtz-Institut f\"ur Strahlen- und Kernphysik (Theorie), 
Universit\"at Bonn, Nu\ss allee 14-16, D-53115 Bonn, Germany 
}

\begin{abstract}
The near-threshold enhancement in the 
$p{\bar p}$ invariant mass spectrum from the 
$B^+{\to}K^+p{\bar p}$ decay reported recently by the BaBar Collaboration is
studied within the J\"ulich $N\bar N$ model. We illustrate that the 
invariant mass  dependence of the $p{\bar p}$ spectrum close to the threshold 
can be reproduced by the final state interactions. This explanation is
in line with our previous analysis of the  $p{\bar p}$ invariant mass 
spectrum from the $J/\Psi{\to}\gamma p{\bar p}$ decay measured by the
BES Collaboration. 
We also comment on a structure found recently in the $\pi^+\pi^-\eta'$ 
mass spectrum of the radiative $J/\Psi$ decay by the BES 
Collaboration.
In particular we argue that one should be rather cautions in 
bringing this structure in connection with the enhancement found
in the $p{\bar p}$ invariant mass spectrum or with the existence
of $N\bar N$ bound states. 
\end{abstract}

\maketitle

A first indication for a near-threshold enhancement in the 
proton--antiproton $p{\bar p}$ invariant mass spectrum 
from the $B^+{\to}K^+p{\bar p}$ 
and ${\bar B^0}{\to}D^0p{\bar p}$ decays were reported by the Belle
Collaboration~\cite{Abe1,Abe2}. Soon afterwards a much more 
significant evidence of a $p{\bar p}$ enhancement,
i.e. with high statistics and high mass resolution, 
was observed by the BES Collaboration~\cite{Bai} in the 
reaction $J/\Psi{\to}\gamma p{\bar p}$.
More recently the Belle Collaboration~\cite{Wang}
found also a near-threshold $p{\bar p}$ enhancement in 
the decays $B^+{\to}\pi^+p{\bar p}$,
$B^+{\to}K^+p{\bar p}$, $B^0{\to}K^0p{\bar p}$ and $B^+{\to}K^{\ast
+}p{\bar p}$, while the CLEO Collaboration detected such
an enhancement in (the unsubtracted) data for 
$\Upsilon (1S) \to \gamma p{\bar p}$ \cite{Cleo}. However,         
in all these cases the results are marred by low statistics. 
Very recently the BaBar Collaboration presented
a new and high statistics measurement of the $B^+{\to}K^+p{\bar p}$ 
decay~\cite{Aubert} confirming the threshold peaking in 
the $p{\bar p}$ invariant mass, see also~\cite{babarweb}. 

The high statistics data by the BES Collaboration triggered a number 
of theoretical speculations where the observed enhancement in the 
invariant $p{\bar p}$ mass spectrum was interpreted as evidence for a 
$p{\bar p}$ bound state or baryonium~\cite{Dover,Shapiro}, 
or for exotic glueball states~\cite{Chua,Rosner}. 
Alternatively, we \cite{Sibirtsev1} but also others 
\cite{Kerbikov,Bugg,Zou,Loiseau} demonstrated that the 
near-threshold enhancement in the $p{\bar p}$ invariant mass 
spectrum from the $J/\Psi{\to}\gamma p{\bar p}$ decay
could be simply due to the final state interactions (FSI) between the 
outgoing proton and antiproton. Specifically, our calculation based 
on the realistic J\"ulich $N{\bar N}$ model~\cite{Hippchen,Mull} and
the one by Loiseau and Wycech \cite{Loiseau}, utilizing the Paris
$N{\bar N}$ model, explicitly confirmed the significance of FSI 
effects estimated in the initial studies~\cite{Kerbikov,Bugg,Zou}
within the effective range approximation. 

In the present paper we want to investigate whether the near-threshold
enhancement in the $p{\bar p}$ invariant mass spectrum, visible in
the high statistics data on the reaction $B^+{\to}K^+p{\bar p}$, 
can likewise be understood in terms of the $p\bar p$ FSI. 
In our study of the $J/\Psi{\to}\gamma p{\bar p}$ decay 
we considered the $p{\bar p}$ FSI interaction in the $^1S_0$ and 
$^3P_0$ partial waves and the $I = 0$ and $I = 1$ isospin channels. 
Other $p\bar p$ $S$- and $P$-waves are ruled out by conservation 
laws for parity, charge-conjugation and total angular momentum 
together with the measured photon angular distribution from the
$J/\Psi{\to}\gamma p{\bar p}$ decay, which agrees with that expected
from the $p{\bar p}$ state being in both $^1S_0$ and $^3P_0$ states.
We found that the mass dependence of the $p{\bar p}$ spectrum close 
to the threshold can be reproduced by the $S$-wave $p{\bar p}$ FSI in 
the isospin $I = 1$ state. 
In case of the $B^+{\to}K^+p{\bar p}$, 
the weak interaction is involved. As a consequence, the selection rules are less
rigid and now other $p\bar p$ $S$- and $P$-waves are
allowed too and could produce FSI effects in the near-threshold 
region. Thus, besides the effects resulting from the 
$^1S_0$ and $^3P_0$ partial waves we explore here also those of
the $^3S_1$ and $^3P_1$ states.

\begin{figure}[t]
\vspace*{-5mm}
\centerline{\hspace*{3mm}\psfig{file=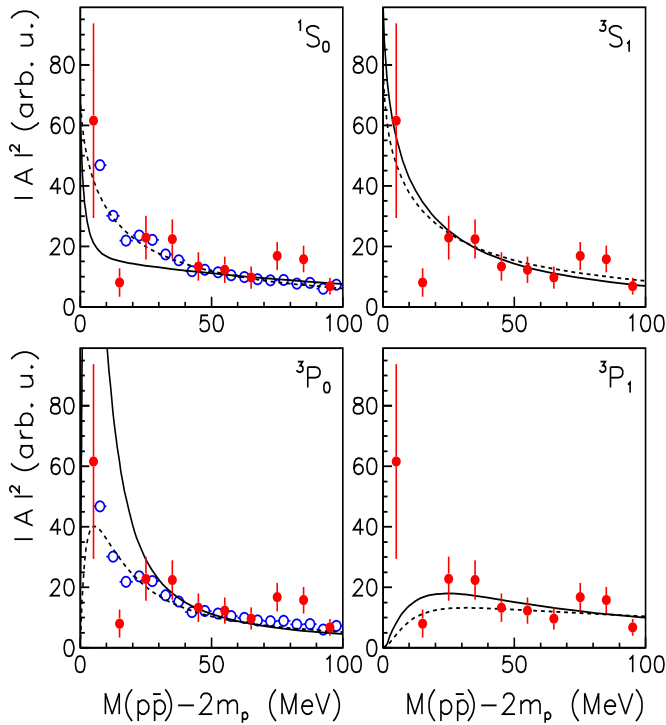,width=9.7cm,height=11.cm}}
\vspace*{-5mm}
\caption{
Invariant $B^+{\to}K^+p{\bar p}$ amplitude $|A|^2$ as a
function of the $p{\bar p}$ mass. The circles represent the experimental
values of $|A|^2$ extracted from the BaBar data \cite{Aubert} via Eq.~(\ref{trans}).
The curves are the scattering amplitude squared ($|T|^2$)
predicted by the $N\bar N$ model A(OBE) \cite{Hippchen}
for the $^1S_0$, $^3S_1$, $^3P_0$,
and $^3P_1$ partial waves and the $I{=}0$ (solid) and $I{=}1$ (dashed) 
channels, respectively. Note that the latter results have been normalized 
to $|A|^2$ at $M(p\bar p){-}2m_p{=}$50 MeV.
For comparison reasons we show also the corresponding results 
extraced from the data on $J/\Psi{\to}\gamma p{\bar p}$ \cite{Bai} 
(open circles). 
}
\label{babar4}
\end{figure}

Like in our earlier paper we utilize the total
spin-averaged (dimensionless) $B^+{\to}K^+p{\bar p}$ reaction amplitude
$A$ and not directly the measured $p\bar p$ invariant mass spectrum, 
because that allows us to get rid of trivial kinematical factors. 
The $B^+{\to}K^+p{\bar p}$ decay rate is given in terms of $A$ 
by~\cite{Byckling}
\begin{eqnarray}
d\Gamma = \frac{|A|^2}{2^9 \pi^5 m_{B^+}^2}\,
\lambda^{1/2}(m_{B^+}^2,M^2,m_{K^+}) \nonumber \\
\times\lambda^{1/2}(M^2,m_p^2,m_p^2)\, dM d\Omega_p\,  d\Omega_K,
\label{spectr}
\end{eqnarray}
where the Kallen function $\lambda$ is defined by
$\lambda (x,y,z)={((x-y-z)^2-4yz})/{4x}\,$,
$M \equiv M(p\bar p)$  is the invariant mass of the $p{\bar p}$ 
system, $\Omega_p$ is the proton angle in that system, 
while $\Omega_K$ is the $K^+$ angle in
the $B^+$ rest frame. After averaging over the spin states and
integrating over the angles, the differential decay rate is
\begin{eqnarray}
\frac{d\Gamma}{dM}=\frac{\lambda^{1/2}(m_{B^+}^2,M^2,m_{K^+})\sqrt{M^2-4m_p^2}}
{2^6 \pi^3 m_{B^+}^2}\,\, |A|^2 \ .
\label{trans}
\end{eqnarray}
We use Eq.~(\ref{trans}) for extracting $|A|^2$ from the data of the
BaBar Collaboration. The corresponding results are shown in 
Fig.~\ref{babar4} by the filled circles. 

We 
assume again the validity of the Watson-Migdal approach for the
treatment of the FSI effect. It suggests that the reaction amplitude
for a production and/or decay reaction that is of short-ranged
nature can be factorized in terms of an elementary (basically
constant) production amplitude and the $p\bar p$ scattering 
amplitude $T$ of the particles in the final state so that
\begin{eqnarray}
A (M(p \bar p)) \approx N \cdot T(M(p \bar p)), 
\label{fsi}
\end{eqnarray}
(cf. Ref. \cite{Sibirtsev1} for further details). 
Thus, we compare the extracted amplitude $|A|^2$ with the 
suitably normalized scattering amplitudes $|T|^2$ that 
result from the J\"ulich $N\bar N$ model. 
The curves shown in Fig.~\ref{babar4} correspond to
the $p{\bar p}$ scattering amplitude squared calculated for 
the $^1S_0$, $^3S_1$, $^3P_0$ and $^3P_1$ partial waves,
where the solid
lines are the results for the isospin $I = 0$ channel, while the
dashed lines are for the $I =1$ channel. 
 
As can be seen from Fig.~\ref{babar4}, the enhancement 
of the near-threshold $p{\bar p}$ invariant mass spectrum from the
$B^+{\to}K^+p{\bar p}$ decay, as observed in the 
new BaBar experiment~\cite{Aubert}, is fully in line with our 
previous results~\cite{Sibirtsev1}. Fig.~\ref{babar4} contains
also the $J/\Psi{\to}\gamma p{\bar p}$ reaction amplitude,
evaluated via 
\begin{eqnarray}
\frac{d\Gamma}{dM}=\frac{(m_{J/\Psi}^2-M^2)\sqrt{M^2-4m_p^2}}
{2^7 \pi^3 m_{J/\Psi}^3}\,\, |A_{J/\Psi}|^2 \ , 
\label{transJ}
\end{eqnarray}
again suitably normalized to 
facilitate an easy comparison of the dependence on the $p\bar p$
invariant mass. 
Obviously the data from both considered decay reactions are in 
reasonable agreement as far as the dependence on $M(p\bar p)$ is 
concerned. 
As already mentioned, while for the $J/\Psi{\to}\gamma p{\bar p}$ 
data the final $p{\bar p}$ system is restricted to the $^1S_0$ and $^3P_0$ 
partial waves in the near-threshold region, due to selection rules 
and the measured photon angular distribution~\cite{Bai}, 
the $B^+{\to}K^+p{\bar p}$ reaction allows also for other partial 
waves. But also here a measurement of the $K^+$ angular distribution 
could clarify whether the $^1S_0$ or $^3P_0$ partial waves 
are responsible for the $p{\bar p}$ enhancement, as for the 
$J/\Psi$, or rather the $^3S_1$, $^3P_1$ or $^1P_1$ states. 
Conservation of the total angular
momentum requires the $K^+$ to be either in a relative $s$ wave
to the $p{\bar p}$ system (for the $^1S_0$ or $^3P_0$ partial waves)
or in a $p$ wave (for $^3S_1$, $^3P_1$ or $^1P_1$).
We note that the invariant amplitude for the $^1P_1$ wave looks very
similar to the one for $^3P_1$.

Recently the CLEO Collaboration published results on the radiative 
decays of the $\Upsilon (1S)$(9460) to the $p\bar p$ system \cite{Cleo}.
Interestingly, also in this reaction one can see an enhancement
in the $p\bar p$ invariant mass spectrum near threshold, cf. Fig. 6
in that paper. The authors
presented also results of a reference measurement for the reaction
$e^+e^-\to \gamma p\bar p$ at the energy $\sqrt{s}$ = 10.56 GeV where 
a similar near-threshold enhancement in the $p\bar p$ mass spectrum
is detected. We do not show 
corresponding results here because the accuracy and the mass resolution 
of those data is too low for allowing a meaningful comparison.
However, we would like to comment on a conclusion drawn in Ref.~\cite{Cleo}.
In order to remove possible continuum background contributions the 
CLEO Collaboration subtracted the (scaled) $e^+e^-\to \gamma p\bar p$ 
mass spectrum from the one measured for the $\Upsilon (1S)$ radiative decay. 
The ``corrected'' $\Upsilon (1S) \to p\bar p$ data do not show an enhancement
in the $p\bar p$ spectrum anymore. We believe that this is not surprising
and, in fact, must be expected if the near-threshold enhancement comes indeed 
from the FSI in the $p\bar p$ system. Then, the same or a similar FSI must be
present in $e^+e^-\to \gamma p\bar p$ as well as in $\Upsilon (1S){\to}\gamma p{\bar p}$
and it must cancel to a large degree in a subtraction like the one 
performed in Ref.~\cite{Cleo}. Accordingly, from our point of view there 
is no contradiction between the results of CLEO and those of the BES 
Collaboration for $J/\Psi{\to}\gamma p{\bar p}$ as suggested in Ref.~\cite{Cleo}. 
Rather the CLEO results even strengthen the conjecture that the 
near-threshold enhancement in the $p\bar p$ spectrum seen in $J/\Psi$,
$B^+$, $\Upsilon (1S)$ etc. decays is due to the $p\bar p$ FSI. 


Next, let us reflect on the present results in view of the earlier 
consideration concering the $J/\Psi$ decays.  
In our work on the $J/\Psi{\to}\gamma p{\bar p}$ spectrum we 
admitted that, because of our poor knowledge of the $N\bar N$ interaction
near threshold and for some other reasons \cite{Sibirtsev1},
explanations for the enhancement other than
final state interactions cannot be ruled out at the present stage.
Specifically, we discussed~\cite{Sibirtsev1} that intermediate
pseudoscalar ($J^{PC}{=}0^{-+}$) meson resonances, for instance 
the $\pi(1800)$ resonance but also the $\eta(1760)$~\cite{PDG}, 
could couple to the $p{\bar p}$ channel and thus could play a role.
In fact, we showed that the presence of these resonances in the 
decay $J/\Psi{\to}\gamma p{\bar p}$ is, in principle, in line with 
the BES data~\cite{Bai} once FSI effects are taken into account.

In this context it is interesting to note that recently the BES
Collaboration reported~\cite{Ablikim} a resonance in the
$J/\Psi{\to}\gamma\pi^+\pi^-\eta^\prime$ mode with mass 
1833.7$\pm$6.1~MeV and width of 67.7$\pm$20.3~MeV obtained by fitting a 
Breit-Wigner function to the $\pi^+\pi^-\eta^\prime$ 
invariant spectrum.
This resonance was denoted as a new $X(1835)$ state, arguing 
that it is not compatible with any of the meson resonance listed in
Ref.~\cite{PDG}.

\begin{figure}[b]
\vspace*{-7mm}
\centerline{\hspace*{8mm}\psfig{file=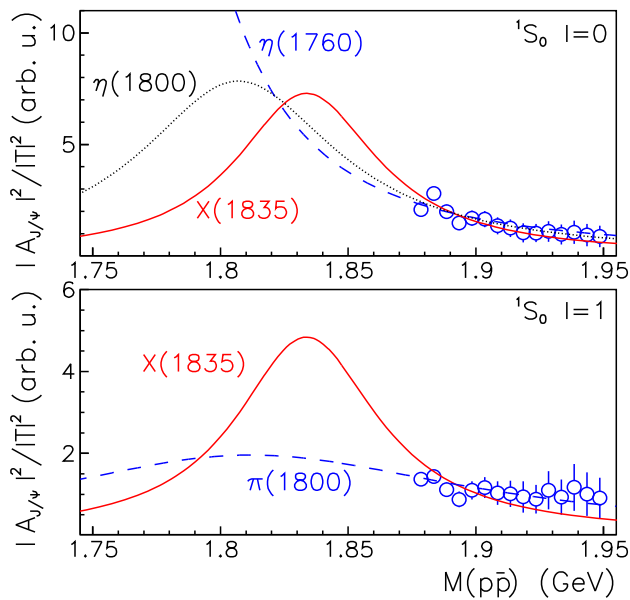,width=9.2cm,height=9.cm}}
\vspace*{-5mm}
\caption{Invariant FSI corrected $J/\Psi{\to}\gamma p{\bar p}$
amplitude $|A_{J/\Psi}|^2$/$|T|^2$ as a 
function of the $p{\bar p}$ mass. The squares show the 
values of $|A_{J/\Psi}|^2$ extracted from the BES data~\cite{Bai} via
Eq.~(\ref{trans}) and divided by the $^1S_0$ scattering amplitude
squared for the $I = 0$ and $I = 1$ isospin states. The lines show the
squared Breit-Wigner amplitudes for the $X(1835)$ \cite{Ablikim}, 
$\eta(1800)$ \cite{Bisello} and $\pi(1800)$ and $\eta(1760)$ \cite{PDG} 
states.}
\label{babar5}
\end{figure}

Following our prescription~\cite{Sibirtsev1} the authors of
Ref.~\cite{Ablikim} re-fitted the $J/\Psi{\to}\gamma p{\bar p}$ 
spectrum including 
a pseudoscalar resonance and the $I = 0$ $^1S_0$ $p{\bar p}$ 
FSI of the J\"ulich $N{\bar N}$ model A(OBE)~\cite{Hippchen}. 
The fit yielded a mass of 1831$\pm$7~MeV and a $\Gamma$ $<$153~MeV  
and led them to the conclusion that the $X(1835)$ properties
as found in $J/\Psi{\to}\gamma\pi^+\pi^-\eta^\prime$ are  
consistent with expectations for the state that produces 
the strong $p{\bar p}$ mass threshold enhancement observed in
the $J/\Psi{\to}\gamma p\bar p$ decay \cite{Ablikim}.

Though the authors of Ref.~\cite{Ablikim} admitted in the
summary that other possible interpretations of the $X(1835)$ 
that have no relation to the observed near-threshold $p{\bar p}$ 
enhancement are not excluded, we think its worthwhile to 
elaborate further on that issue. 
Dividing the average total $J/\Psi{\to}\gamma p{\bar p}$ 
reaction amplitude $|A_{J/\Psi}|^2$, which is related to the
differential decay rate via Eq.~(\ref{transJ}),
by the $p{\bar p}$ scattering amplitude 
$|T|^2$~\cite{Hippchen,Sibirtsev1} one can easily obtain the
FSI corrected data as a function of the $p{\bar p}$ invariant mass
$M(p{\bar p})$, cf. Fig.~\ref{babar5}. 
Here we use the $^1S_0$ scattering amplitudes in the $I = 0$ and $I = 1$ 
isospin states.
 
In our experience the data do not allow to fix uniquely the
resonance properties if the mass, width and the strength of the 
resonant contribution are unknown. 
In order to illustrate that we show the
squared Breit-Wigner amplitudes for the $X(1835)$, the $\eta(1760)$ and
$\pi(1800)$ with their properties given by the BES
Collaboration~\cite{Ablikim} and the PDG~\cite{PDG}, respectively. We
only vary the coupling strength of the $0^{-+}$ state to the $p{\bar p}$
channel in order to reproduce the BES data~\cite{Bai}. 
Since the $X(1835)$ might be a $I^G(J^{PC}){=}0^+(0^{-+})$ 
resonance \cite{Ablikim} one can certainly speculate whether it is the
same object as the $\eta(1760)$ listed by the PDG~\cite{PDG}. In fact,
the $\eta(1760)$ was also established in radiative $J/\Psi$ decays, 
namely in the reaction $J/\Psi\to \gamma \rho\rho$ \cite{Bisello}. 
While the PDG cites only an averaged value, a glance into the original
paper \cite{Bisello} makes clear that the mass as well as the width 
of the $\eta(1760)$ could not be reliably established from the data. 
Indeed, one of the six solutions with comparable $\chi^2$ given in
Ref. \cite{Bisello} (in Table IV) yields a mass and width of 
1807$\pm$10 MeV and 94$\pm$12 MeV, which is not that far away from the
values obtained by the BES Collaboration.  We show also results based 
on the above resonance parameters in Fig.~\ref{babar5}. It is obvious
that there is practically no difference between the $X(1835)$ of
Ref.~\cite{Ablikim} and the $\eta(1800)$ of Ref.~\cite{Bisello} as
far as the description of the $p\bar p$ invariant mass spectrum
is concerned. 

Anyway, the properties of the $X(1835)$ are indeed consistent with the
measured near-threshold enhancement in the $p{\bar p}$ spectrum of
the reaction $J/\Psi{\to}\gamma p{\bar p}$. On the other hand, one
has to concede that this enhancement as such does not provide any 
reliable additional support for the existence of the $X(1835)$ 
resonance, and likewise not for the existence of $p{\bar p}$ bound states
or baryonia. With regard to the latter we want to remind the reader
that our model calculations \cite{Sibirtsev1} as well as those of 
Loiseau and Wycech \cite{Loiseau} are able to reproduce the $p{\bar p}$ 
spectrum. However, while the $N\bar N$ model used in \cite{Loiseau}
generates a near-threshold bound state (at $E \approx -5 -i50$~MeV) in
the relevant $^{11}S_0$ state no such state is present in our $^{31}S_0$
amplitude (which describes the data equally well) and the one in the 
$^{11}S_0$ (at $E \approx -104 -i413$~MeV \cite{Mis}) is too wide to have 
an influence in the physical region.
 
We believe that it will be rather difficult to establish experimentally
a direct connection between the $X(1835)$ and the $p\bar p$ system.
On the other hand, it would be still interesting to investigate whether
this resonance is also visible in $p\bar p$ annihilation. The $X(1835)$
could be searched for in reactions like $p\bar p \to \pi^+\pi^-X$, 
$X\to \pi^+\pi^-\eta'$, etc. \cite{Klempt}. Measurements to get 
more information on these issues could be performed using the 
PANDA detector at the future FAIR project. 

In this context let us also remind the reader that standard quark-model
calculations like those in Ref.~\cite{Stanley,Isgur,Metsch} do predict radial
excitations of the $\eta$, $\eta'$ around the $\eta$(1800) mass. Thus,
a very conventional interpretation of the structure found by the BES
Collaboration should be also taken into consideration before 
speculating excessively on exotic explanations. 

In summary, we have analyzed the near-threshold enhancement in the 
$p{\bar p}$ invariant mass spectrum from the $B{\to}Kp{\bar p}$ 
decay reported recently by the BaBar Collaboration within the 
J\"ulich $N\bar N$ model. Our study shows that the mass 
dependence of the $p{\bar p}$ spectrum close to the threshold 
can be reproduced by the final state interaction. This explanation is
in line with our previous investigation of the $p{\bar p}$ invariant mass 
spectrum from the $J/\Psi{\to}\gamma p{\bar p}$ decay measured by the
BES Collaboration.

\acknowledgments{
We thank Mathew Graham for providing us with the most recent BaBar data.
This work was partially supported by the 
DFG (SFB/TR 16, ``Subnuclear Structure of Matter''), by
the EU Integrated Infrastructure Initiative Had{\-}ron Physics Project 
(contract no. RII3-CT-2004-506078)
and by Department of Energy under contract  no.
SURA-06-C0452.}


\begin{thebibliography}{99}
\bibitem{Mis} Note that there is a misprint in the binding energy
        given in Ref. \cite{Sibirtsev1}.
\end{thebibliography}
\end{document}